\documentclass[12pt]{iopart}
\usepackage{iopams}  
\usepackage{harvard}
\usepackage{bm}
\usepackage{amssymb}
\newcommand{\R}{\Bbb R}

\newcommand{\scrif}{{\mathcal{I}^{+}}}

\newcommand{\const}{\mathrm{const}}
\newcommand{\conj}{\mathrm{conjugate}}

\newcommand{\z}{{\mathfrak z}}

\begin{document}

\title[Center-of-mass ambiguity for BMS charges]{Center-of-mass ambiguity\\
for Bondi--Metzner--Sachs charges}

\author{Adam D. Helfer}

\address{Department of Mathematics, and Department of Physics \& Astronomy,\\ 
University of Missouri, Columbia MO 65211 USA}
\ead{helfera@missouri.edu}

\begin{abstract}
Dray and Streubel proposed a definition of angular momentum in general relativity based on `Bondi--Metzner--Sachs (BMS) charges'.  I show here that the natural definition of center of mass in this program has an infinite-dimensional ambiguity.  (This seems not to have been noticed before because previous 
work
has tacitly carried over a special-relativistic assumption.)  
A related point is that the natural definition of spin in this context is translation-, but not supertranslation-, invariant.
\end{abstract}


\section{Introduction}

The problem of giving a satisfactory treatment of angular momentum for Bondi--Sachs space--times is well-known \cite{Szabados2009}.  In one sense, the difficulty is that angular momentum is an origin-dependent quantity, and there are arguments showing that, in the presence of gravitational radiation, it is impossible to find a preferred Poincar\'e-covariant asymptotic Minkowski space of origins \cite{ADH2007}.  
Indeed, there is not even a preferred asymptotic Poincar\'e group, but rather a much larger object, the Bondi--Metzner--Sachs (BMS) group.

While the Poincar\'e group is a collection of motions {\em preserving} an important physical structure (Minkowski space), the BMS group is a rather weak, infinite-dimensional, object, introduced to {\em compensate} for lack of structure  
(or, if you prefer, to encompass many different allowable structures).  Nevertheless, there is a strong formal parallel between the Poincar\'e and BMS groups, and a main branch of the work on angular momentum is based on trying to exploit this.  Dray and Streubel proposed such a definition, which has been much used \cite{DS1984,Dray1985}.\footnote{The idea of using the BMS motions to give angular momentum goes back at least to Winicour's {\em linkages} \cite{WT1965}, and forms the basis of the Ashtekar--Streubel {\em fluxes} \cite{AS1981}.  Interestingly, Dray and Streubel were influenced in an essential way by Penrose's twistor approach \cite{Penrose1982}, although the twistor treatment of angular momentum is {\em not} based the BMS group at all \cite{ADH2007}.  Recently, some authors have suggested expanding the Dray--Streubel ideas
to an even larger class of asymptotic motions than the BMS group (bringing in `superrotations' and `superboosts'); 
see \cite{FN2017} and references therein.}

In this parallel, the weakness of the BMS group leads also to the replacement of the Minkowski four-dimensional space of origins with the infinite-dimensional space of `cuts' of $\scrif$.
%\footnote{Recall that for a Bondi--Sachs space--time, future null infinity $\scrif\simeq\{ (u,\theta ,\phi )\mid u\in\R,\ (\theta,\phi )\in S^2\}$ where I am abusing notation by identifying the polar coordinates $(\theta ,\phi)$ with a point on $S^2$; this point is an asymptotic null direction and $u$ is a Bondi retarded time coordinate.  The BMS group is the semidirect product of the Lorentz group and the {\em supertranslations}, which have the form $u\mapsto u+\alpha (\theta ,\phi )$ for suitably smooth $\alpha$.  The projection $\scrif\to S^2$ is invariant.  A {\em cut} is a section $u=\z (\theta,\phi )$ of this, and can be thought of as an instant of retarded time.}
%
In fact, because we are dealing with a situation in which degrees of freedom may escape the system in radiation, the Dray--Streubel angular momentum is in effect a function $\bm{M_{ab}}(\z_{\rm act},\z_{\rm pas})$ of two cuts:  in Szabados's terminology, the {\em active} cut $\z_{\rm act}$, representing the retarded time at which we wish to know how much angular momentum remains in the system; and the {\em passive} cut $\z_{\rm pas}$, which serves as the origin about which the angular momentum is measured.\footnote{This usage does not really correspond to the notions of active and passive transformations, but no harm will be done if we are aware of this.  The terms are helpful in making clear a distinction which will be central here.}  This angular momentum takes values in a space of asymptotically constant skew Lorentz tensors.
There is a change-of-origin formula, which describes how this varies with $\z_{\rm pas}$, which will be the core of this paper.

In special relativity, the tensorial character of the angular momentum is critical for comparing relatively rotated or boosted systems.  However, when we consider a single system, to interpret its angular momentum we almost invariably pass to its rest frame, where we may read off its spin and center of mass.  
(Recall that the center of mass is unique up to translations along the energy--momentum.)
We may seek to do the same for the Dray--Streubel angular momentum at any cut $\z _{\rm act}$.

I will point out here two concerns which come up when we do this.  These go to the question of whether the charges are being interpreted correctly as angular momentum.  
It is possible that a way of resolving these will be found, but at the moment one, especially, of the issues appears to lead to difficulties of
interpretation.

The first, lesser, issue attaches to extracting the spin (and likely has been observed by others).  In special relativity, in a system's rest frame, the spin comprises the space-space components of the relativistic angular momentum, and these are translation-invariant.  One might hope that in the general-relativistic case, the space-space components of the Dray--Streubel angular momentum would be {\em supertranslation}-invariant in the rest frame; this 
would be what the hypothesized correspondence between the Poincar\'e and BMS treatments would suggest.  However, this is not true:  the space-space components in the general-relativistic case are only translation-invariant.  This may be counted as a disappointment, in the sense that it contravenes the strongest results we might have hoped for with the BMS-based approaches; but it does not obviously violate any physical principles.

The situation for the center of mass is more of a concern.  We should not generally expect that the center of mass would differ from the cut $\z_{\rm act}$ simply by a translation.  (For example, in a non-Minkowski stationary space--time, there is a clear candidate for a one-parameter family of cuts, related by time-translation, which should be considered the center of mass.  But $\z_{\rm act}$ can be any cut, and so it need not be related to these by a translation.)  We may seek the center of mass by looking for cuts $\z_{\rm pas}$ for which the time-space components of $\bm{M_{ab}}(\z_{\rm act},\z_{\rm pas})$ vanish.
What we shall find is that it is indeed possible to do this, but there is an infinite-dimensional freedom in doing so, and the different solutions are in a physically significant sense incompatible (they are relatively supertranslated).

This result can in fact be read off from known formulas.  I think it has not been noticed before, or at any rate not appreciated, because workers have carried over special-relativistic expectations.  Most commonly, what has been computed is the quantity $\bm{M_{ab}}(\z_{\rm act},\z_{\rm act})$, where the origin has been taken to be the same as the active cut, and the time-space components of this, divided by the mass, have been taken to be the center of mass (relative, again, to $\z_{\rm act}$).  This always produces a center of mass which is a translation of $\z_{\rm act}$ --- but as I pointed out above, this is not generally the best answer, even for a stationary space--time.  While of course it has been appreciated that one {\em could} make other choices of the reference cut $\z_{\rm pas}$, and while the formulas for making such supertranslations form an integral part of the theory, there seems to have been no shift in point of view, that one should allow the supertranslational freedom into the determination of center of mass.

The result above, that what appears to be the most natural way to try to define the center of mass leads to an infinite-dimensional, mutually incompatible, family of candidates, 
calls into question the interpretation of the time-space components of $\bm{M_{ab}}(\z_{\rm act},\z_{\rm pas})$ as the center of mass.

Of course, in the stationary case, one can in a well-motivated way simply restrict the construction to the `good' cuts, and then no difficulty arises.  However, the situation for the non-stationary case is troublesome.  One might try to resolve the problem by finding some preferred set of cuts to work with.  But then (apart from the fact that this is almost diametrically opposite from the motivating principle of BMS invariance), one risks losing a more basic feature:  universality.  If the definition of angular momentum cannot be made at any cut $\z _{\rm act}$, or if the set of admissible origins $\z_{\rm pas}$ depends on the details of the space--time under consideration, the very concept of angular momentum depends on the space--time, and one cannot meaningfully compare the angular momenta of different space--times.  (Usually one expects that, if one makes a choice of identification of the $\scrif$s of two different space--times, one should be able to compare their energy--momenta and angular momenta, just as, in special relativity, once one fixes the Poincar\'e motion relating one system to another, one can compare their energy--momenta and angular momenta.)

These points emphasize the need to go beyond the formal mathematical correspondence between the Poincar\'e and BMS groups, and develop a detailed physical interpretation of the structures occurring in the BMS-charge approach to angular momentum.  

{\em Organization, conventions and background.}
The next section contains preliminaries, the one after 
gives the details of the argument, and the final section contains discussion.  This paper assumes at least a passing familiarity with the Newman--Penrose spin-coefficient formalism at $\scrif$, for which see \cite{PR1986}, whose conventions I follow.  (These are also compatible with the papers of Dray and Streubel.)
I will denote quantities taking values in the space of asymptotically constant tensors \cite{ADH2014} by boldface, so they are easily distinguishable.  The speed of light is unity, and Newton's constant is $G$.

%A few comments about the technicalities are in order.  Interestingly, the result on the center of mass needs almost no technical work --- it really just depends on understanding the translations and supertranslations.  

\section{Preliminaries}

For a Bondi--Sachs space--time, future null infinity $\scrif$ has canonically the structure of a bundle of affine lines over $S^2$; I will write $\scrif \simeq \{ (u,\theta ,\phi )\mid u\in\R,\ (\theta ,\phi )\in S^2\}$ (abusing notation by representing a point on $S^2$ by its polar coordinates).  We may interpret the sphere here as the set of asymptotic future null directions; one calls $u$ a {\em Bondi retarded time parameter}.  The BMS group is the semi-direct product of the (proper, orthochronous) Lorentz group with the {\em supertranslations}, the maps of the form $(u,\theta ,\phi )\mapsto (u+\alpha (\theta ,\phi), \theta ,\phi )$ for suitably smooth $\alpha$.  

It turns out that the supertranslations with only $l=0$, $l=1$ multipole components form an invariant subgroup; these may be identified with the translations.  However, there is no invariant sense of a `translation-free supertranslation'.  Because of this, all (suitably smooth) sections $u=\z (\theta ,\phi )$ of the fibration $\scrif\to S^2$ are BMS-equivalent.  They are called {\em cuts} of $\scrif$, and interpreted as instants of retarded time, in the sense that they mark off portions $u\leq\z (\theta ,\phi)$ of $\scrif$ from which radiation prior to $\z$ has escaped.

The detailed study of the radiative regime is commonly done in terms of the Newman--Penrose spin-coefficient formalism at $\scrif$.  It would be out of place to give an exposition of that here.  However, only a relatively small number of very technical features are used in this paper, and they are set out explicitly.  

In fact, the main result, the ambiguity in the center of mass, depends only on the properties of the translations and supertranslations given above, and the fact that the Dray--Streubel change-of-origin formula reduces to the special-relativistic one for translations.

The other result, that the spin is sensitive to supertranslations, does involve some more technicalities, having to do with the multipole properties of a quantity called the {\em mass aspect} and the appearance of the vector field generating rotations in the Newman--Penrose formalism.  But, accepting these points, the reader will find that the argument reduces to a straightforward one involving familiar multipole decompositions.

The main points the reader will need about the Newman--Penrose formalism at $\scrif$ are these.  It is based on a complex null tetrad adapted to the Bondi coordinate system, with $m^a$ an antiholomorphic tangent to the $u=\const$ cuts.  (One convenient choice would be $m^a=2^{-1/2}(\partial_\theta +i\csc\theta\partial_\phi )$.)
All tensors are reduced, using this tetrad, to `spin-weighted' quantities on the sphere (elements of appropriate line bundles).  The operator $\eth$ (with complex conjugate $\eth'$) is a natural derivative operator on these quantities, in the $m^a$ direction (and so reducing to $m^a\nabla_a$ on scalars).  

Relevant properties of the BMS motions are that the translations are the supertranslations with $\eth^2\alpha =0$, and that a Lorentz generator $cm^a+{\overline c}{\overline m}^a$ on the sphere corresponds to a rotation iff $\eth c +\eth'\overline{c}=0$.

\section{Spin and center of mass in the Dray--Streubel approach}

It will be enough to consider the angular momentum for a fixed choice of `active' cut $\z_{\rm act}$.  As usual, I will suppose the Bondi coordinate system is chosen so that this is a cut of constant coordinate $u=u_{\rm act}$, and I will also assume the time-axis of the Bondi system is chosen to align with the energy--momentum at this cut.

Let the `passive' cut, the origin-cut relative to which the angular momentum is measured, be $u=\z_{\rm pas}$; this 
%is 
need not be a slice of constant $u$.  Let $cm^a+\overline{c}{\overline m}^a$ be the vector field on the sphere determining a Lorentz generator
$\bm{\Lambda_{ab}}(c)$
(so $c$ is a spin-weight $-1$ quantity satisfying $\eth' c=0$).  Then the BMS charge
representing the component 
$-(1/2)\bm{\Lambda^{ab}}(c) \bm{M_{ab}}(\z_{\rm act},\z_{\rm pas})$
of the angular momentum
about $u=\z _{\rm pas}$ may be written
\begin{eqnarray}
  Q &=&Q_0+Q_{\rm super}\, ,
\end{eqnarray}
where
\begin{eqnarray}
  Q_0 &=&    \frac{-1}{8\pi G}\oint\{ 
   (\psi _1+\sigma\eth\overline\sigma +(1/2)\eth{\sigma\overline\sigma })c\}
   +\conj 
\end{eqnarray}
is independent of $\z_{\rm pas}$ and
\begin{eqnarray} \label{Qf}
\fl  Q_{\rm super} 
   &=&\frac{-1}{8\pi G}\oint\{
     [ 2(c\eth+{\overline c}\eth' )  (\z_{\rm pas} -u_{\rm act}) - (\eth c+\eth'{\overline c})(u_{\rm act}-\z_{\rm pas})]\Re (\psi_2+
           \sigma\dot{\overline\sigma} ) \}\qquad
\end{eqnarray}
contains the effects of the supertranslation of $\z_{\rm pas}$ relative to $u_{\rm act}$.  (The integrals are evaluated at $u=u_{\rm act}$, and the standard area-form on the sphere is understood.)

The quantity $\psi_2+\sigma\dot{\overline\sigma}$ is called the {\em mass aspect,} and 
one of its properties will be important: its real part has, in our choice of frame (aligned with the Bondi--Sachs energy--momentum), no $l=1$ multipole component.
(In general, the imaginary part of the mass aspect has no $l=0$ or $l=1$ components; in our frame, the mass aspect has no $l=1$ component.)

\subsection{Rotations and spin}

Consider first the case of rotations.  For these, 
the quantity $\eth c+\eth'{\overline c}=0$.  In this case, if $\z_{\rm pas}-u_{\rm act}$ is a translation, we will have $Q_{\rm super}=0$.  This is because
the quantity $(c\eth+{\overline c}\eth')(\z_{\rm pas}-u_{\rm act})$ will be a pure $l=1$ multipole, and this will vanish when integrated against $\Re(\psi_2+\sigma\dot{\overline\sigma})$.  
The space-space components of the Dray--Steubel angular momentum are thus invariant under translations.  

On the other hand, unless $\Re(\psi_2+\sigma\dot{\overline\sigma})$ is constant over the sphere (which would most notably occur in a stationary space--time), 
these space-space components will not generally be invariant under supertranslations.  This is because the only functions on the sphere invariant under all rotations are the constants.
If the real part of the mass aspect is not constant, we can find some axis about which it has some real multipole with a non-trivial $m\not=0$ contribution.  
(Here $m$ is the degree of the multipole, the number of periods about the axis.)
Then we may take $cm^a+{\overline c}{\overline m}^a $ to generate a rotation about that axis, and adjust $\z_{\rm pas}$ so that  $(c\eth+{\overline c}\eth' )(\z_{\rm pas} -u_{\rm act})$ 
is the same multipole; the integral will then be positive.
In particular, if $\Re (\psi_2+\sigma\dot{\overline\sigma})$ has any non-zero multipole for $l\geq 2$, we can arrange $Q_{\rm super}\not=0$.

\subsection{Center of mass}

Now let us look at the center of mass.  As I pointed out above, we should not generally expect the center of mass, evaluated at a cut $\z_{\rm act}$, to be related to this cut simply by a translation.  It should be supertranslated.  To allow for this, the natural thing to do is to define the center of mass to be given by those cuts $\z_{\rm pas}$ for which the time-space components of $\bm{M_{ab}}(\z_{\rm act},\z_{\rm pas})$ vanish.

There is an infinite-dimensional family of such cuts.  To see this, suppose we start with {\em any} cut $\z$, and consider a {\em translation} of $\z$ by $\tau$.  We then have the change-of-origin formula
\begin{eqnarray}
  \bm{M_{ab}}(\z_{\rm act},\z +\tau) 
      &=&\bm{M_{ab}}(\z_{\rm act},\z)
        +\bm{P_a\tau_b} -\bm{P_b\tau_a}\, ,
\end{eqnarray}
where $\bm{P_a}$ is the Bondi--Sachs energy--momentum and $\bm{\tau^a}$ represents $\tau$ in the space of asymptotically constant vectors.  This formula is the same as the special-relativistic one.  Assuming we are not in the Minkowski-space case, 
we have $\bm{P_a}\not=0$, and there will be a unique choice of $\tau$ (modulo constants, which represent translations along $\bm{P_a}$) making the time-space components vanish.  Denote this $\tau_{\rm comp}=\tau _{\rm comp}(\z)$, the translation complementary to $\z$.  

What we have shown is that, for {\em any} cut $\z$, the passive cut
\begin{eqnarray}
  \z_{\rm pas} =\z +\tau _{\rm comp}
\end{eqnarray}
lies on the center of mass.   Notice that if two cuts $\z_1$, $\z_2$ differ by a supertranslation, then the centers of mass we get $\z_{1,{\rm pas}}$, $\z_{2,{\rm pas}}$ must also differ by a supertranslation,\footnote{And if $\z_1$ and $\z_2$ differ by a translation, then $\z_{1,{\rm pas}}$, $\z_{2,{\rm pas}}$ must be equal modulo addition of a constant.}
which reflects a significant physical incompatibility.
This shows that there are as many different centers of mass as there are `translation-free supertranslations', or more formally that the centers of mass may be identified with the quotient
Supertranslations/Translations.  Any element of this quotient has a unique (up to constants, representing translations along $\bm{P_a}$) representative which is a center of mass, and inversely any center of mass determines an element of the quotient.

\section{Discussion}

I considered here what one would take to be the most natural approach to defining center of mass within the Dray--Streubel program.  This approach seemed on its face to hold the possibility of significant physical insight, for it suggested that one might start with a cut $\z_{\rm act}$ without any special properties, yet recover from that a center of mass $\z_{\rm pas}$ selected by the physics of the situation.  That this does not appear to work, and that we are instead presented with an infinite-dimensional set of disparate candidates, calls us to reexamine our premises.

The main assumptions were that, in the Bondi frame aligned with the energy--momentum, the time-space components of $\bm{M_{ab}}(\z_{\rm act},\z_{\rm pas})$  are the first mass-moments with respect to the translations, and that all choices of $\z_{\rm pas}$ are admissible.
As long as we keep these,
we are led to the conclusion that each element in the quotient Supertranslations/Translations has a unique (modulo translations along $\bm{P_a}$) representative whose first mass-moments vanish, and it is hard to avoid interpreting this as a candidate center of mass.

Possibly one could look for some ancillary condition to impose, to restrict the class of passive cuts $\z_{\rm pas}$ which ought to be physically considered.  This 
would be a 
step away from the BMS invariance which motivates the approach,
but it is reasonable to consider such an adjustment.  
A restriction would have to be carefully chosen, for making one
which depended on the particulars of the space--time would likely make it harder to compare angular momenta in different situations, and lessen the quantities' usefulness.  (This is the universality referred to in the introduction.)
Closely related but not quite identical concerns have led a number of authors to look for a preferred class of {\em active} cuts; see \cite{AKN2009,GM2014,KQ2016,KNQ2020} and references therein.

The supertranslation-noninvariance of the spin does not seem at all as problematic, in and of itself.  But that the spin is tied to the mass-moments in the Lorentz-covariant quantity  $\bm{M_{ab}}(\z_{\rm act},\z_{\rm pas})$ suggests that there are further points we do not understand.

Finally, I should mention that the considerations of this paper apply also to the Ashtekar--Streubel fluxes \cite{AS1981}, since those can be viewed as the limit of differences of BMS charges \cite{Dray1985}.  The verification of this is straightforward.

\section*{Acknowledgment}

It is a pleasure to thank Tevian Dray for helpful comments.  Of course, the perspective here, and any errors, are mine.

\section*{References}

\bibliographystyle{jphysicsB}

%\begin{Harvard}

%\bibliography{../ReferencesZ}

%\end{Harvard}

\end{document}